\documentclass[twocolumn,superscriptaddress]{revtex4-1}
\usepackage{graphicx}
\usepackage[T1]{fontenc}
\usepackage[latin1]{inputenc}
\usepackage{graphicx}
\usepackage{amsmath}
\usepackage{color}

\begin{document}
\title{Properties of patchy colloidal particles close to a surface: a Monte Carlo and density functional study}

\author{Nicoletta Gnan}
\email{nicoletta.gnan@roma1.infn.it}
\affiliation{Dipartimento di Fisica, Universit\`{a} di Roma ``La Sapienza'', 
Piazzale A. Moro $2$, $00185$ Roma, Italy}

\author{Daniel de las Heras}
\affiliation{Theoretische Physik II, Physikalisches Institut, Universit\"at Bayreuth, $D-95440$ Bayreuth, Germany}

\author{Jos\'{e} Maria Tavares}
\affiliation{Instituto Superior de Engenharia de Lisboa, Rua Conselheiro Em\'{\i}dio Navarro, P-1590-062 Lisbon, Portugal, and Centro de F\'{i}sica Te\'orica e Computacional da Universidade de Lisboa, Avenida Professor Gama Pinto 2, P-1649-003, Lisbon, Portugal}

\author{Margarida M. Telo da Gama}
\affiliation{Departamento de F\'{\i}sica, Faculdade de Ci\^encias da Universidade de Lisboa, Campo Grande, P-1749-016, Lisbon, Portugal, and Centro de F\'{i}sica Te\'orica e Computacional da Universidade de Lisboa, Avenida Professor Gama Pinto 2, P-1649-003, Lisbon, Portugal}

\author{Francesco Sciortino}
\affiliation{Dipartimento di Fisica, Universit\`{a} di Roma ``La Sapienza'', 
Piazzale A. Moro $2$, $00185$ Roma, Italy}

%\date{\today}
\begin{abstract}
We investigate the behavior of a patchy particle model close to a hard-wall via Monte Carlo simulation and density functional theory (DFT). Two DFT approaches, based on the 
homogeneous and inhomogeneous versions of Wertheim's first order perturbation theory for the association free energy are used. We evaluate, by simulation and theory, the 
equilibrium bulk phase diagram of the fluid and analyze the surface properties for two isochores, one of which is close to the liquid side of the gas-liquid coexistence curve. 
We find that the density profile near the wall crosses over from a typical high-temperature adsorption profile to a low-temperature desorption one, for the isochore close to 
coexistence. We relate this behavior to the properties of the bulk network liquid and find that the theoretical descriptions are reasonably accurate in this regime.  At very 
low temperatures, however, an almost fully bonded network is formed, and the simulations reveal a second adsorption regime which is not captured by DFT. We trace this failure 
to the neglect of orientational correlations of the particles, which are found to exhibit surface induced orientational order in this regime.

\end{abstract}

\maketitle

\section{Introduction \label{introduction}}

Colloidal particles with patterned surfaces -- better known as patchy 
particles~\cite{GlotzerNanoL4,BianchiPCCP13,PawarMRC31,SciortinoCurr2011} -- 
have been studied extensively in recent years owing to their ability to self-
assemble in a rich number of cluster, gel, glassy and crystalline phases.
Understanding how the surface pattern influences the self-assembly mechanism 
is crucial to a bottom-up strategy for designing new materials where the 
desired macroscopic behavior 
is encoded in the microscopic properties of the building-blocks\cite{Flavio}. Patchy 
particles represent a valuable model system for investigating and 
understanding the behavior of more complex constituents such as amphiphilic 
molecules, colloidal clays, proteins and DNA nano-
assemblies~\cite{SciortinoPCCP12,RosenthalJCP134,RuzickaNM10, DoyePCCP9, 
GlotzerNM9}. New concepts, 
as equilibrium gel \cite{SciortinoCurr2011}, optimal network 
density~\cite{DeMicheleJCP125} and empty liquid have arisen from the study of 
the phase diagrams~\cite{delasHerasJCP134,delasHerasSM7} of patchy particles 
with limited valence~\cite{bianchi2008,BianchiPRL2006}. These models emphasize 
the role of the number of bonds between the particles in determining the 
equilibrium as well as the static and dynamic behavior of the system in and 
out of equilibrium~\cite{RussoPRL2010}. 

Homogeneous patchy particle fluids are described satisfactorily by Wertheim's 
first order thermodynamic perturbation 
theory~\cite{wertheim1,wertheim2,wertheim3,wertheim4} that provides an 
expression for the free energy per particle with $n_p$ patches, which are 
treated independently.
Less known are the properties of patchy particle fluids in confined 
geometries. Understanding the behavior of patchy particles close to surfaces 
has direct impact on a number of different applications which require patchy 
particles to self-assemble in confined geometries. We recall for instance the 
templated self-assembly technique~\cite{Bishop2009} where confined geometries 
are used to orient bulk structures or to induce the formation of novel 
morphologies. Moreover recent studies have focused on effective forces between 
colloids generated by confined critical patchy particles, aiming to control colloids stability~\cite{GnanSM2012}.
Finally such systems may be studied with small-angle neutron 
scattering~\cite{SteitzEPL67} and atomic force microscopy \cite{TibergCOCIS4} 
and thus a quantitative microscopic description of patchy particles in 
confined geometries is highly desirable. 

The description of confined associating fluids has been addressed in the past. 
Density functional theory (DFT) based on a perturbation of the inhomogeneous 
hard-sphere fluid was used to describe the structure of inhomogeneous 
associating fluids. 
An early example is the work of Segura et al.~\cite{Segura97} for particles 
with four patches close to a hard wall where the weighted density 
approximation (WDA) of Tarazona ~\cite{Tarazona1,Tarazona2} was combined with 
Wertheim's theory to obtain a DFT description of associating fluids. In the 
original DFT formulation both homogeneous and inhomogeneous versions of 
Wertheim's first order perturbation theory were used to account for the 
association free energy of the particles and it was found that the homogeneous 
theory - where the law of mass action is identical to that of the bulk system 
with the density replaced by the averaged local density - yields satisfactory 
results over the whole range of parameters. By contrast, the inhomogeneous 
version of Wertheim's theory was found to overestimate badly the layering of 
the particles near the wall. The reasons for this discrepancy are not fully 
understood but they appear to be related to the difficulty of implementing 
Wertheim's law of mass action in the inhomogeneous system. This approach was 
extended to mixtures of associating and neutral equi-sized hard-
spheres~\cite{Segura99}. 

A second approach for the same model - particles with four patches - was 
developed by Yu and Wu~\cite{yu:7094}. They employed a modified fundamental 
measure theory (FMT)~\cite{Rosenfeld} based on Rosenfeld's functional for 
inhomogeneous hard-sphere fluids and an inhomogeneous version of Wertheim's 
free energy for association. Yu and Wu~\cite{yu:7094} employed an 
inhomegeneous version of the law of mass action which was shown to give 
comparable results to the homogeneous theory of Segura 
et al. for intermediate densities. Yu and Wu's theory, however, is more 
accurate at the highest densities and can be extended to binary mixtures of 
particles with different sizes. More importantly, it reveals that the 
inhomogeneous version of Wertheim's first order perturbation theory depends 
crucially on the generalization of the law of mass action for inhomogeneous 
patchy particle systems.  

Both theories were applied to systems where bonding is not fully developed, 
i.e. at temperatures that are not too low, and the relation between surface 
and bulk properties has not been investigated. In this paper we focus on the 
self-assembly of patchy particles with $n_p=3$ at a planar hard wall, and 
extend our calculations to the region where a fully bonded optimal network 
develops. We have simulated the system by fixing the density and scanning the 
temperature down to very low temperatures. We compare the simulation results 
to the results of two density functional theories: a homogeneous WDA approach 
similar to that 
of Segura et al. and the inhomogeneous FMT approach of Yu and Wu, which were 
shown to be equivalent for hard-spheres in the range of densities considered 
here. For the three-patch particle model the theories give similar results at 
moderate to high temperatures, by contrast to the low-temperature regime, 
relevant to the formation of arrested states, where both theories break down. 
We trace this failure to the assumption of independent patches and stress the 
need for the development of density functionals that account for the 
directionality of the bonding, which plays a crucial role in the low-
temperature regime.

The manuscript is organized as follows: in Sec. II we describe the model and 
the simulation techniques. Moreover, a brief comparison of the results of 
Wertheim's theory and the simulation results for the bulk fluid are presented 
and discussed. In Sec. III we describe the two density functional approaches 
and in Sec. IV  we discuss their validity and limitations, with emphasis on 
the low-temperature regime.

\section{Background}\label{theory}

\subsection{Model and simulation methods}
\begin{figure}[h]
	\centerline{\includegraphics[width=.9\linewidth]{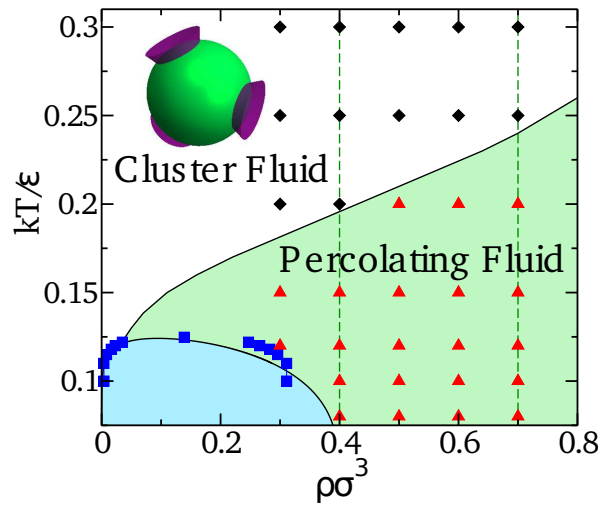}}
	\caption{Phase diagram of patchy particles with valence $n_p=3$. Symbols are results from MC simulations. Squares depict the gas-liquid coexistence line. Diamonds indicate cluster fluid states ($p_b<0.5$), while triangles indicate percolating fluid states. Solid lines are the results from Wertheim's theory for the coexistence line and from Flory-Stockmayer theory for the percolation line. 
	Dashed lines indicate the isochores investigated in this study. Patchy particles are modeled as hard spheres with $3$ equidistant bonding sites (patches) on the particle equator (inset).}
	\label{fig1}
\end{figure}

\begin{figure}[h]
	\centerline{\includegraphics[width=.9\linewidth]{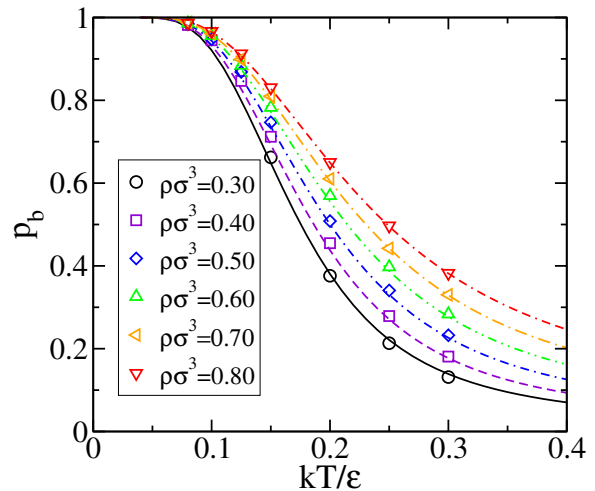}}
	\caption{Comparison of the bonding probability $p_b$ from Monte Carlo simulations (symbols) and Wertheim's theory (Eq.~\ref{eq:pbRelation}, lines).}
	\label{fig2}
\end{figure}

Patchy colloidal particles are modeled as hard spheres (HS) of diameter $\sigma$ with $n_p=3$ equidistant bonding sites on the particle equator (see the inset of Fig~\ref{fig1}). The interaction between two patches $V_{ij}$ belonging to particles $i$ and $j$ is given by the Kern-Frenkel potential~\cite{kern9882}:

\begin{equation}
	V_{ij}=V_{SW}(|\vec r_{ij}|)G(\hat r_{ij},\hat r_i,\hat r_j),
\end{equation}

\noindent where $\vec r_{ij}$ is the vector between the centers of particles $i$ and $j$, and $\hat r_i$ is the unit vector from the center of particle $i$ to the center of one patch on its surface. $V_{SW}$ is a square well potential,

\begin{equation}
	V_{SW}(|\vec r_{ij}|)=\left\lbrace
 	\begin{array}{l}
 		\infty\quad \text{ if}\quad |\vec r_{ij}|<\sigma,\\
 		-\varepsilon \quad \text{ if}\quad \sigma\le |\vec r_{ij}|\le \sigma+\delta,\\	 
		 0  \quad \text{ otherwise}, 
  	\end{array}
  	\right.
\end{equation}

\noindent and $G$ is the angular part:

\begin{equation}
	G(\hat r_{ij},\hat r_i,\hat r_j)=\left\lbrace
 	\begin{array}{l}

	   1 \text{ if}\quad\left\lbrace
		\begin{array}{l}
  	    	\hat r_{ij}\cdot\hat r_i>\cos(\theta_{max}),\\
       		-\hat r_{ij}\cdot\hat r_j>\cos(\theta_{max}),\\
       	\end{array}
       \right.\\

	   0  \quad \text{ otherwise}.
  	\end{array}
  	\right.
\end{equation} 

\noindent The interaction energy between sites $\varepsilon$ sets the energy scale. The spatial range $\delta$ and the angle $\theta_{max}$ control the volume available for bonding, $v_b$, which is:
\begin{equation}
	v_b=\frac{\pi\sigma^3}3\left[(1+\delta/\sigma)^3-1\right]\left[1-\cos(\theta_{max})\right]^2.
\end{equation} 

\noindent We fix the parameters $\delta=0.119\sigma$ and $\cos \theta_{max}=0.895$, fulfilling the single bond per patch condition assumed in Wertheim's first order perturbation theory.

We perform Gibbs ensemble Monte Carlo (GEMC)~\cite{PanagiotopoulusMP61} simulations to locate the gas-liquid coexistence line and grand-canonical Monte Carlo (GCMC) simulations~\cite{Frenkelbook} to estimate the critical point. In the GEMC method, a MC step consists on average of $4000$ roto-translation attempts, $400$ particle swap attempts and one volume change. About 300 particles in a volume $V=2880 \sigma^3$  were simulated. In the GCMC method, we consider boxes with $L=6\sigma$ to $L=14\sigma$, and a MC step consists of $500$  roto-translation attempts and one insertion/deletion attempt. For the largest box, the number of particles fluctuates between zero and 800.
We scan the chemical potential $\mu$ and temperature $T$ to locate the region where the system exhibits large fluctuations in the number of particles $N$ and in the energy $E$, which signal the presence of a critical point. The appropriate combination of these fluctuations, at the critical point, follows the order parameter distribution of the Ising universality class~\cite{WildingJPCM9,VinkJCP121}.  

The surface properties are investigated by canonical ensemble Monte Carlo (MC) simulations, at different temperatures and densities, by introducing a surface in the middle of the simulation box and periodic boundary conditions. The surface is modeled by a planar hard wall acting on the particles through a hard-core repulsion, i.e., the interaction between the particles and the surface is purely entropic. 
The density profiles close to the wall are calculated at two bulk densities, at temperatures $T$ above and below the percolation line. 

\subsection{Wertheim Theory}

Within Wertheim's theory, the free energy for a homogeneous system is the sum of two contributions. The free energy of the hard-sphere reference system and the bonding contribution $F_{bond}(\rho,T)$ which arises from considering certain graphs in the Mayer expansion~\cite{HansenMcDonald}. For the present single-component model, the bonding contribution can be expressed in terms of the bonding probability $p_b$ (fraction of bonded sites) as

\begin{equation}\label{eq:Bond_Free_En}
	\beta F_{bond}/N=n_p \ln(1-p_b)+\frac{1}{2}n_p p_b,
\end{equation}

\noindent where $\beta=1/kT$ with $k$ the Boltzmann constant. Assuming that all sites have the same probability of bonding, Wertheim's theory predicts that $p_b$ is determined by the law of mass action:

\begin{equation}\label{eq:pbRelation}
	\frac{p_b}{(1-p_b)^2}=\rho n_p \Delta,
\end{equation}

\noindent where $\Delta$ is the equilibrium constant of the ``reaction'' (bonding) between two
patches and $\rho$ is the density. To evaluate $\Delta$ for the present model, 
we assume that the radial distribution function of the reference HS fluid $g_{HS}(r)$ is approximated by \cite{NezbedaMolPhys69}:

\begin{equation}\label{eq:grhs}
	g_{HS}(r)=(A_0+A_1)+A_1 (r/\sigma -1),
\end{equation}

\noindent where 

\begin{eqnarray}
 A_0 &=& \frac{1-0.5\eta}{(1-\eta)^3}+\frac{4.5\eta (1+\eta)}{(1-\eta)^3},\\
 A_1 &=& \frac{-4.5\eta (1+\eta)}{(1-\eta)^3}.
 \end{eqnarray}

As a result, 
\begin{eqnarray}\label{eq:Delta}
	\Delta &=& 4\pi\chi^2\left[\frac{(1+\delta)^3-1}{3}A_0+\frac{(1+\delta)^4-1}{4}A_1\right]\\\nonumber
	& & \times \left[\exp(\beta \varepsilon)-1\right].
\end{eqnarray}

\noindent where $\eta =\frac{\pi\rho\sigma^3}{6}$ is the packing fraction and $\chi =0.5(1-cos(\theta_{max}))$ is the fraction of surface covered by the patch. Note that the linear approximation of Eq.~\ref{eq:grhs} is highly accurate in the relevant $r$ range, i.e.  within the well of the square-well potential.
     
The phase diagram is then calculated straightforwardly. It includes a gas-liquid first order phase transition that ends at a critical point at low $T$ and $\rho$. The coexisting homogeneous phases have different densities and fractions of unbonded sites. The percolation line is calculated using the Flory-Stockmayer (FS) theory of polymerization \cite{flory1,flory2,stock1} which gives for the percolation threshold, 

\begin{equation}
	p_b=\frac1{n_p-1}.
\end{equation}

\subsection{Bulk behavior}

Fig.~\ref{fig1} shows the phase diagram obtained via GCMC and GEMC simulations and compares it with the theoretical results from Wertheim's theory.
The figure also shows the FS percolation line. The high density phase is always percolated, in the sense that there is a non-zero probability of finding an infinite cluster that contains almost all the particles. The percolation line ({\it i. e.} the line that separates percolated from non-percolated states) intercepts the binodal on the low-density phase, 
near the critical point.
As shown in Fig.~\ref{fig1}, the percolation line starts on the left of the critical point and the temperature increases monotonically with the density. In Fig. \ref{fig2} we plot the bonding probability as a function of temperature for different densities. The agreement between Wertheim's theory and Monte Carlo simulations is quite good both for the phase diagram and for the bonding probability.

\section{Density functional theory}
As usual in density functional theory, we split the Helmholtz free-energy functional into the ideal and excess parts:

\begin{equation}
	F[\rho(\vec r)]=\mathcal{F}_{id}[\rho(\vec r)]+\mathcal{F}_{exc}[\rho(\vec r)], 
\end{equation} 

\noindent where $\rho(\vec r)$ is the number density. Here and in what follows, we assume that the single-particle distribution function $\rho(\vec r)$ depends on the spatial but not on the orientational coordinates. Note that the arrangement of patches on the equator breaks the rotational symmetry of the particles, and therefore states with a preferred orientation of the particles cannot be ruled out. In fact, this is a crude approximation that will break down at low temperatures, as we will discuss later. 

The ideal part is given by: 

\begin{equation}\label{eq:F_id} 
 \beta \mathcal{F}_{id}[\rho(\vec r)]=\int d^3r \rho(\vec r)\left[\ln(\rho(\vec r){\cal V})-1\right], 
\end{equation} 

\noindent where ${\cal V}$ is the thermal volume. The integral is over the volume $V$. 

$\mathcal{F}_{exc}$ contains the excluded volume interactions between HS and the bonding free energy due to bond formation between the particles. We have used two different approximations for $\mathcal{F}_{exc}$: a modified version~\cite{Kim} of the local weighted density approximation (WDA) introduced by Segura et al.~\cite{Segura97} and the fundamental-measure theory (FMT) for associating fluids developed by Yu and Wu~\cite{yu:7094}. 

In the WDA method, the HS and the association term in the free-energy per particle are evaluted for a homogeneous system at the same weighted density. The latter is calculated for a fluid of HS \cite{Kim}. The Carnahan-Starling~\cite{HansenMcDonald} approximation is employed for the HS contribution, while the association free energy is given by Wertheim's first-order perturbation theory for a homogeneous system(Eq.~\ref{eq:Bond_Free_En}).

In the FMT proposed by Yu and Wu~\cite{yu:7094}, Wertheim's free-energy functional for the inhomogeneous system is used and thus the two contributions are treated separately. The HS reference fluid is described by the Rosenfeld FMT approach while the association contribution is given by an appropriate inhomogeneous Wertheim's term. A detailed description of both methods is found in appendices \ref{LWDA} and~\ref{FMT} respectively. 

\section{Results}
\begin{figure*}
	\centerline{\includegraphics[width=1\linewidth]{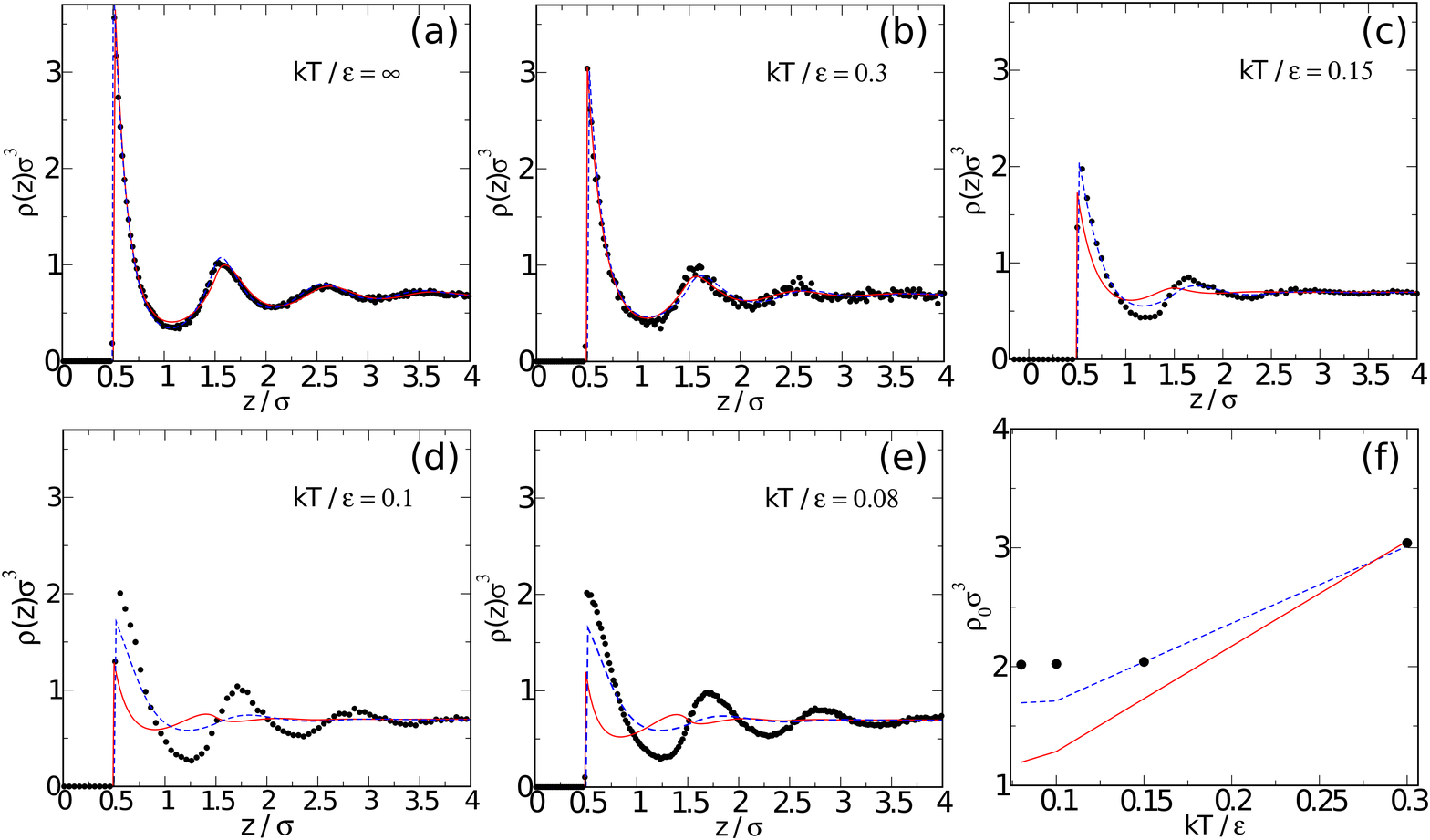}}
	\caption{Density profiles as a function of the distance from the wall at different temperatures for $\rho_{b}\sigma^3=0.70$: (a) $kT/\varepsilon\rightarrow\infty$, (b) $kT/\varepsilon=0.30$, (c) $kT/\varepsilon=0.15$, (d) $kT/\varepsilon=0.10$, and (e) $kT/\varepsilon=0.08$. The full circles are Monte Carlo simulation results. The lines are the results from density functional 	theory: FMT (red solid line) and WDA (dashed blue line). (f) Contact density as a function of temperature.}
	\label{fig3}
\end{figure*}

\begin{figure*}
	\centerline{\includegraphics[width=.80\linewidth]{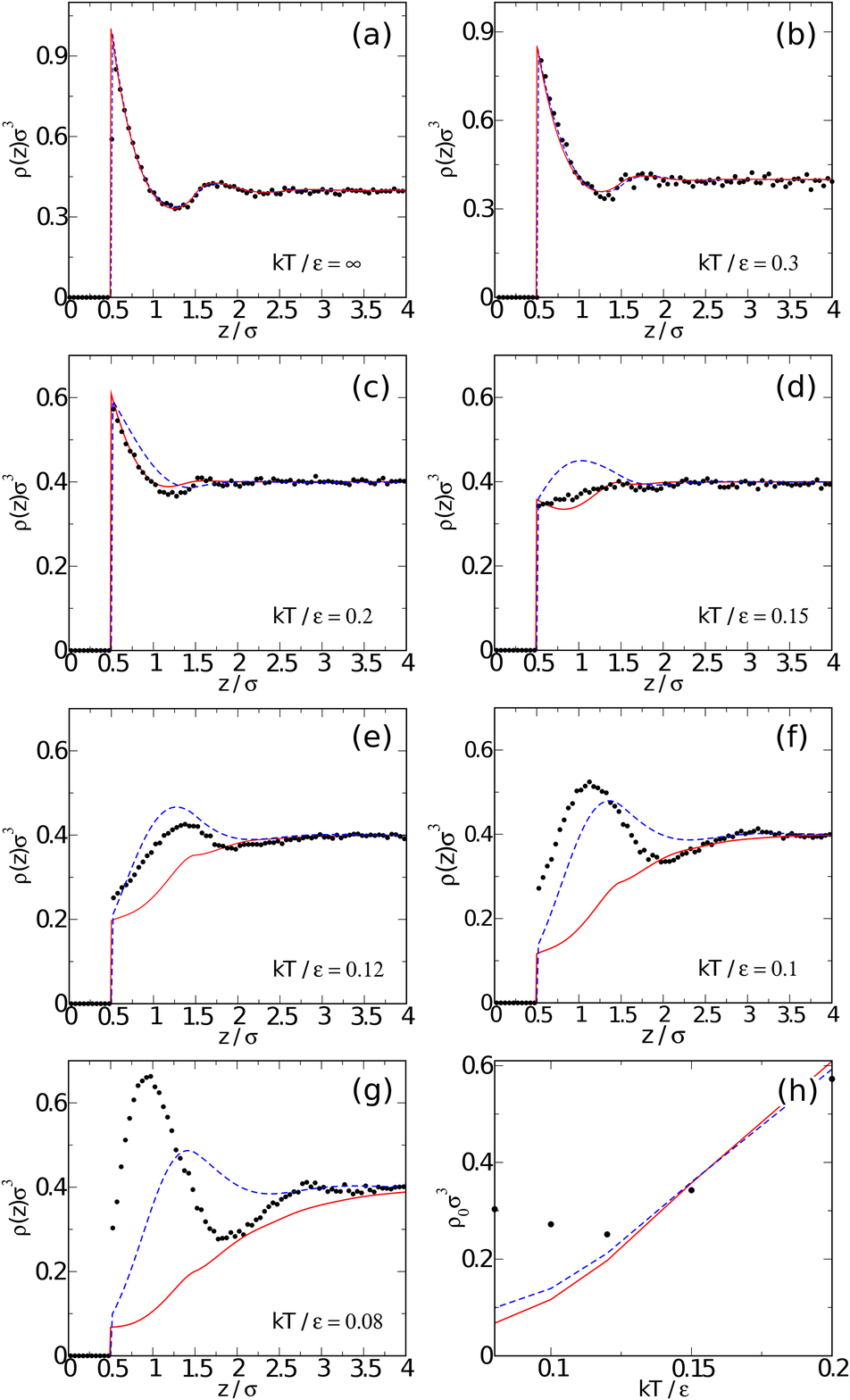}}
	\caption{Density profiles as a function of the distance from the wall at different temperatures for $\rho_{b}\sigma^3=0.40$: (a) $T\rightarrow\infty$, (b) $kT/\varepsilon=0.30$, (c) $kT/\varepsilon=0.20$, (d) $kT/\varepsilon=0.15$, (e) $kT/\varepsilon=0.12$, (f) $kT/\varepsilon=0.10$, and (g) $kT/\varepsilon=0.08$. The full circles are Monte Carlo simulation results. Lines are density functional theory results: FMT (red solid line) and WDA (dashed blue line). (h) Contact density as a function of temperature.}
	\label{fig4}
\end{figure*}

We focus on the surface properties of the fluid at two bulk densities ($\rho_{b}\sigma^3=0.70$ and $\rho_{b}\sigma^3=0.40$) in contact with a neutral hard-wall, and consider different temperatures (see Fig.~\ref{fig1}). The main results are reported in Fig.~\ref{fig3} and Fig.~\ref{fig4}.

The density profiles calculated from MC simulations for $\rho_{b}\sigma^3=0.70$, are plotted together with FMT and WDA theoretical results in Fig.~\ref{fig3}.
The limiting case of HS is also included for reference (panel (a) of Fig.~\ref{fig3}). Panels (b)-(d) of Fig.~\ref{fig3} show the results for three different reduced temperatures $T$ above and below the percolation line. The main difference between the associating fluid (finite temperature) and the reference HS fluid ($T\rightarrow\infty$) is the reduction in the adsorption of particles near the wall. This effect can be understood as follows: at very high temperatures almost all the particles are fully unbonded, and there is a strong adsorption of particles near the wall due to entropic reasons (gain in configurational entropy). As we decrease the temperature, the level of association between particles increases. The probability of bonding at distances of order $\sigma$ from the wall is lower than at larger distances (the wall is neutral). The contribution to the bonding free energy of the particles near the wall is lower than the contribution due to the other particles, resulting in a reduction of the adsorption of patchy particles at the wall when compared to the adsorption of hard-spheres (see the contact density, {\it i. e.} the value of the density at $z/\sigma=0.5$, in panel (f) of Fig. \ref{fig3}).

The WDA provides a fairly good description of the density profiles down to $kT/\varepsilon=0.15$, i.e. well inside the percolation region (see Fig. \ref{fig1}). At this temperature WDA seems slightly more accurate than FMT, providing a good estimate of the contact density. The agreement is lost at $kT/\varepsilon=0.10$ and below (Fig.~\ref{fig3} (d) and (e)) where both theories underestimate the contact density and do not account for the strong fluid layering on increasing the distance from the wall. 
At this density, we find that patchy particles are adsorbed on the wall at all $T$ (although the adsorption decreases as the temperature decreases). No changes with further cooling are expected at lower temperatures since $p_b(kT/\varepsilon=0.08)\approx 0.98$, meaning that the liquid has reached an almost fully bonded configuration and thus the structural properties become essentially $T$-independent~\cite{ZacPRL94,SciortinoCurr2011}. 
 
In Fig.~\ref{fig4}, panels (a)-(g), we plot the density profiles for $\rho_b\sigma^3=0.40$. Panel (a) illustrates the HS fluid corresponding to $T\rightarrow\infty$.
At $kT/\varepsilon=0.3$ (b) and $kT/\varepsilon=0.2$ (c) both, FMT and WDA, are in reasonable agreement with the MC simulation results. FMT yields slightly better results than WDA for $kT/\varepsilon=0.2$. Again, the two density profiles indicate an adsorption of particles near the wall. The adsorption, due to the excluded volume, is small compared to the adsorption of HS (Fig.~\ref{fig4} (a)). 
%Those particles near the wall have a reduced probability of bonding (the wall is neutral), giving rise to a desorption of particles at small distances from the wall (compared with a pure fluid of HS or, alternatively, a fluid of patchy particles at infinite temperature). Nevertheless, at this high temperature, the excluded volume effects dominate and the probability of finding one particle near the wall is higher than in bulk. 
Fig.~\ref{fig4} (d) illustrates the system at  $kT/\varepsilon=0.15$ (slightly below the percolation threshold). The MC and FMT density profiles are almost uniform. Patchy particles are slightly desorbed from the wall, indicating the cancellation between the effects due to excluded volume and association. At this temperature WDA predicts a desorption of particles very close to the wall, but it also predicts the formation of a more or less well-defined layer of particles at approximately $z/\sigma=1$ from the wall. Panels (e)-(g) illustrate the low temperatures: $kT/\varepsilon=0.12$ (e), $kT/\varepsilon=0.10$ (f), and $kT/\varepsilon=0.08$ (g). At these $T$ the association of particles is very high (see the fraction of unbonded sites in Fig.~\ref{fig6}), and the energy of bonding dominates the behaviour of the system. As a result, FMT predicts a strong desorption of particles from the wall. However, the simulation shows the opposite behavior; there is a well-defined layer of particles near the wall. The layer grows and approaches the wall as the temperature decreases. This desorption-adsorption crossover is also reflected in the contact density depicted in panel (h) of Fig.~\ref{fig4}. WDA predicts the presence of a layer of particles at approximately $z\approx \sigma$ for low $T$. Nevertheless, the peak heights do not vary significantly with $T$, by contrast to the simulation results. 
 
The overall $T$ dependence of the contact density at $\rho_{b}\sigma^3=0.40$ is significantly different from the system at $\rho_{b}\sigma^3=0.70$. Indeed, for $\rho_{b}\sigma^3=0.40$, the contact density changes continuously from the HS adsorption limit to the low $T$ desorption, while in the  $\rho_{b}\sigma^3=0.70$ system only adsorption is present. 
To rationalize this behavior we recall the thermodynamics of the bulk system, and in particular the gas-liquid coexistence. In limited valence systems, the liquid side of the gas-liquid coexistence is almost vertical in the $T-\rho$ plane. The associated density provides a quantification of the so-called optimal network density, \textit{i.e.} the density at which particles in the liquid (actually a gel at low $T$) are not stressed. At the same time, the small value of the density of the gas-phase indicates that the coexisting pressure is rather small. The contact density, at a hard-wall, is a direct measure of the pressure in the system. Hence, close to the coexisting liquid branch, the contact density decreases as $T$ decreases, resulting in the reported desorption. On increasing $\rho$ at constant low $T$, the formation of an extended network of bonds causes the increase of stresses in the system and the pressure increases. This results in a significantly large value of the contact density, driving the adsorption phenomenon. 

\begin{figure}
	\centerline{\includegraphics[width=.95\linewidth]{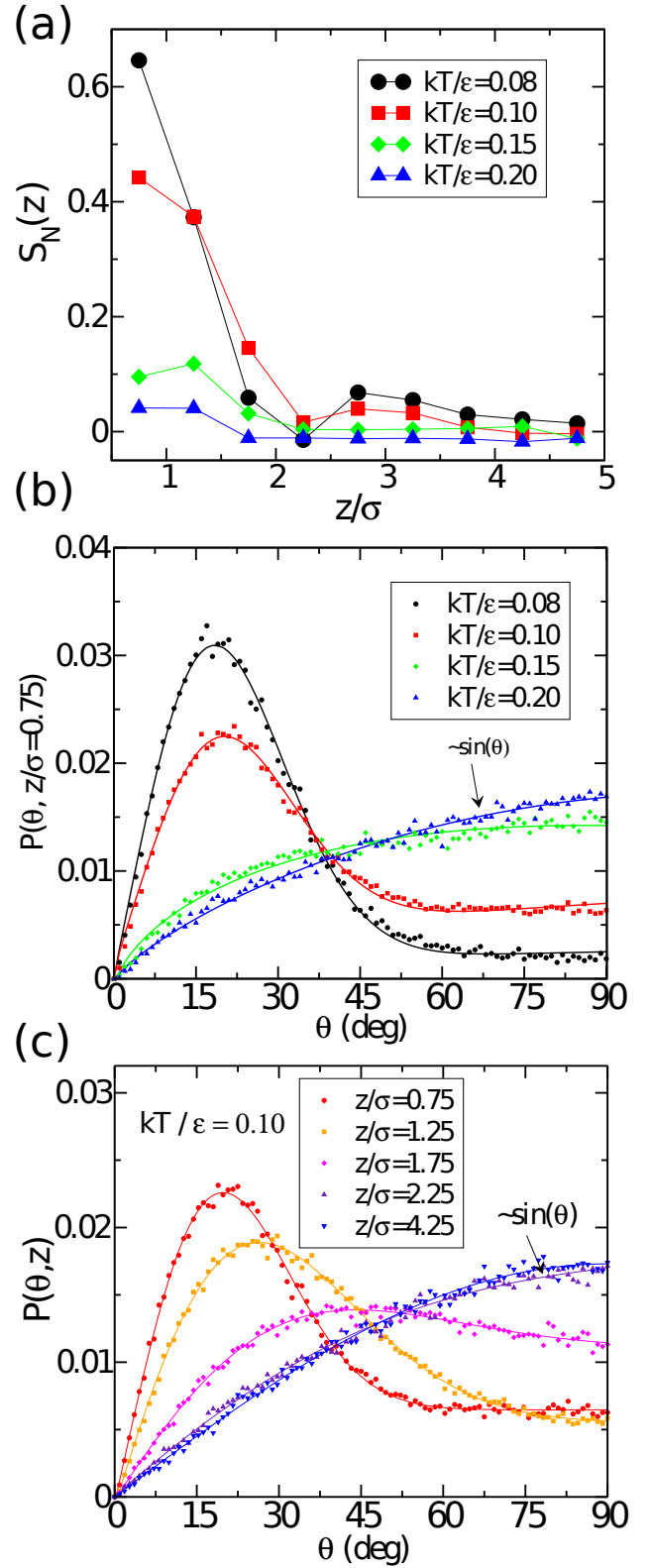}} 
	\caption{(a) Orientational order parameter $S_N(z)$ near the wall extracted from MC canonical simulation of patchy particles for different $T$ and $\rho_b\sigma^3=0.40$. (b) Probability distributions $P(\theta,z)$ as a function of $\theta$ for $z=0.75\sigma$ at the same $T$ and $\rho$ as in panel (a). (c) Probability distributions $P(\theta,z)$ as a function of the distance from the wall $z$ evaluated at $kT/\varepsilon=0.1$ and $\rho\sigma^3=0.4$. $\theta$ is evaluated at different distances from the wall by dividing the simulation box into ``slices'' of size $L \times  L \times \Delta z$ where $\Delta z=0.5\sigma$. Notice that when orientational isotropy is restored (high $T$), $P(\theta) \sim \sin (\theta)$. Lines are guides to the eye.}
	\label{fig5}
\end{figure}

To pin down the origin of the discrepancies between the DFT results and MC simulations which build up on cooling, we plot in Fig.~\ref{fig5} (a) the uniaxial order parameter profile $S_N(z)$ for the system at $\rho_{b}\sigma^3=0.40$. This is defined as $S_N(z)=\int\,d\hat\Omega\,h(\hat\Omega,z)$ $P_2(\cos(\theta))$ where $h(\hat\Omega,z)$ is the orientational distribution function at distance $z$ from the wall, $P_2(\cos(\theta))$ is the second Legendre polynomial and $\theta$ is the angle between the unit vector $\vec{u}$ normal to the wall and the unit vector $\vec{p}$ normal to the plane containing the patches. $S_N(z)$ provides information on the orientation of particles as a function of the distance $z$ from the wall. As shown in Fig.~\ref{fig5}, $S_N(z)$ grows significantly close to the wall on cooling. The value of $S_N(z)$ signals a cross-over from an isotropic fluid to a nematic-like phase near the wall. To highlight the particles orientation close to the wall, we show in  Fig.~\ref{fig5}(b) the probability distributions $P(\theta,z)$, confirming that at low $T$ the particles are oriented near the wall, with the plane containing the patches parallel to the wall. This geometry maximizes the bonding probability by moving the patches away from the neutral wall. The average orientational angle close to the wall decreases continuously with $T$ and it should approaches $\theta=0$ for perfect order. We note indeed that the finite bonding volume allows for a flexibility in the orientation of the particles, contributing to a small but non-zero average angle even in highly bonded  conditions close to the wall. At high $T$ the particles are randomly oriented (see Fig. \ref{fig5} (a) and (b)) and $P(\theta,z) \sim \sin(\theta)$. Finally we note that far from the wall (Fig.~\ref{fig5} (c)) particles undergo a continuous change from a state with preferred orientation  to an isotropic one.

We stress that both DFT approaches neglect the orientational order that develops in the system at low T, since the free-energy depends on the number density but not on the orientation of the particles. 

\begin{figure}
	\centerline{\includegraphics[width=.95\linewidth]{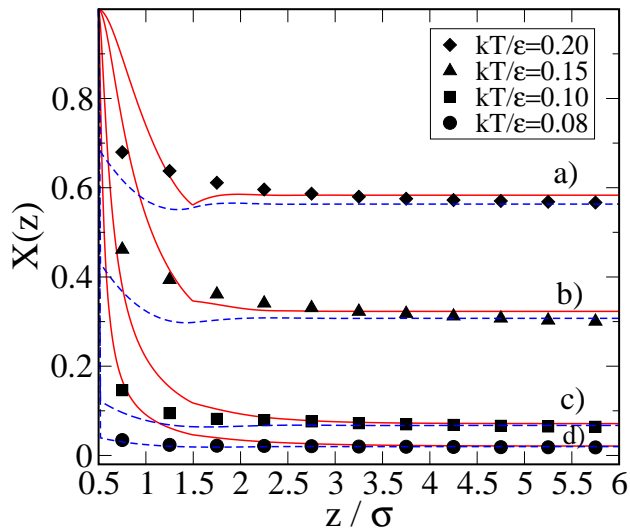}}
	\caption{Fraction of unbonded sites as a function of the normal distance from the wall at different $T$s and $\rho_b\sigma^3=0.40$. Symbols are results from MC simulation. Solid lines are results from the FMT-inhomogeneous Wertheim theory (eq. \ref{eq:inhomog_X}). Dashed lines are obtained from the WDA-homogeneous Wertheim theory (eq. \ref{eq:pbRelation}).}
	\label{fig6}
\end{figure}

Finally we comment on the difference between the results of the two DFT approaches at low $T$ for $\rho_b\sigma^3=0.40$. Even though the agreement with the numerical results is poor 
in both cases, WDA describes qualitatively the layering of the particles near the wall at low $T$ while this feature is absent in the FMT results. Such layering is related - in part - to the orientation of the particles (which is neglected in both DFTs) in order to minimize the fraction of unbonded sites $X(z)$ (i.e. $1-p_b$) of particles near the wall.
Fig.~\ref{fig6} shows $X(z)$ along $z$ as predicted by FMT (solid lines) and WDA (dashed lines).  Symbols are the results from MC simulations. At low $T$, $X(z)$ obtained from the homogeneous Wertheim theory (WDA) exhibits a trend similar to the MC results, while $X(z)$ calculated using the inhomogeneous Wertheim theory (FMT) increases sharply near the wall. 
The WDA $X(z)$ follows the density profile, increasing over the bulk value when $\rho(z)> \rho_b$, which appears to describe more accurately the $z$ dependence of $X$. A somewhat related but more drastic failure of an inhomogeneous version of Wertheim theory was reported by Segura et. al.~\cite{Segura97}.

\section{Concluding remarks}
In this work we have studied the properties of a fluid of patchy particles with $n_p=3$ patches. We have located the gas-liquid coexistence line, the critical point and the percolation line, providing a full characterization of the thermodynamics and structure of the fluid phases. We have also shown that, for this patchy model, Wertheim's theory for homogeneous fluids is accurate. We have then studied the surface properties of this fluid in contact with a neutral hard-wall using MC simulation and two DFT approaches.  

We have investigated the behavior of the system at two different densities, one close to the liquid branch of the coexistence curve and one about 70 per cent higher, for several $T$,
covering the structural change of the system from a monomer solution to an almost fully bonded network state.
We have shown that at low $\rho_b$, the wall adsorbs particles at high $T$ which desorb as $T$ is lowered. The physical mechanism responsible for the adsorption-desorption cross-over is understood in terms of the proximity to the gas-liquid coexistence curve. Indeed, at low $T$, close to the liquid branch, the liquid coexists with a gas at very low density and the pressure is small. Since the contact density at the hard wall is a measure of this pressure, at small $\rho_b$ desorption must occur. On increasing $T$ the pressure increases and for 
$T$ above the percolation threshold the particles are adsorbed at the wall.
Such crossover is not observed at the higher $\rho_b$ where the density at the wall is always larger than the bulk density. Indeed, even at low $T$, the price to pay for the formation of a distorted network leads to an increase of the pressure and hence to a large contact density, even when $p_b \rightarrow 1$.

Not surprisingly, the description using density functional theory is consistent with simulations and with the results reported in Ref. \cite{Segura99} and \cite{yu:7094} at high and intermediate $T$, but it fails at low $T$. We have traced this failure to the inadequacy of describing the orientational degrees of freedom of the particles. Indeed, close to the wall, the particles are oriented in such a way that the plane containing the bonding sites is almost parallel to the wall. It appears that the orientational degrees of freedom and an appropriate coupling to the density profile need to be taken into account in future work in order to describe the structure of associating fluids (and gels) close to a hard wall at low temperatures.

\section{Acknowledgements}
NG and FS acknowledge support from ERC-$226207$-PATCHYCOLLOIDS. They also thank A. Parola for discussions and suggestions.
%\textcolor{red}{add here your acknowledgements}
DdlH acknowledges the support from the Spanish Ministry of Education (contract no. EX2009-0121) and Programme of Activities (Comunidad de Madrid, Spain) MODELICO-CM/S2009ESP-1691.
JMT and MMTG acknowledge financial support from
the Foundation of the Faculty of Sciences of the University of Lisbon and FCT,
under Contracts nos. PEst-OE/FIS/UI0618/2011 and PTDC/FIS/098254/2008, and 
from the FP7 IRSES Marie-Curie grant
PIRSES-GA-2010-269181.

\appendix
\section{Weighted density functional}\label{LWDA}
The weighted density functional approach is based on the idea that the free energy of the inhomogeneous system, characterized by the single-particle density $\rho(\vec{r})$, may be written in terms of the free energy of a homogeneous fluid with an effective density that is evaluated through an appropriate averaging procedure. The approach was proposed by Tarazona\cite{Tarazona2} and was used and modified by several authors \cite{Curtin,Rosenfeld,Kierlik}.

The excess free energy functional of the inhomogeneous fluid is written as 
\begin{equation}\label{eq:F_ex}
	\mathcal{F}_{exc}[\rho]=\int d\vec{r} \rho(\vec{r}) f[\rho;\vec{r}],
\end{equation}
where $f[\rho;\vec{r}]$ is the local excess free energy functional per particle. $f[\rho;\vec{r}]$ can be written as a function of the density $\bar{\rho}$, a functional of the single-particle density, which satisfies
\begin{equation}\label{eq:mapping}
	f[\rho;\vec{r}]=f(\bar{\rho}(\vec{r})),
\end{equation}
where $f(\bar{\rho}(\vec{r}))$ is the local free-energy density of the homogeneous system and $\bar{\rho}(\vec{r})$ is the weighted density defined by: 
\begin{equation}\label{eq:rho_bar}
	\bar{\rho}(\vec{r})=\int d\vec{r}'\rho(\vec{r}')\omega(\vec{r}-\vec{r}';\tilde{\rho}(\vec{r})).
\end{equation}

In Eq. \ref{eq:rho_bar} $\tilde{\rho}(\vec{r})$ is the difference between the density and its weighted counterpart and $\omega$ is the weight function satisfying the constraint
\begin{equation}
	\int d\vec{r}' \omega(\vec{r}-\vec{r}',\tilde{\rho}(\vec{r}))=1.
\end{equation}
The weight function, which encodes the non-local character of the functional, is related via non linear differential equations to the direct correlation function of the inhomogeneous fluid. An approximation to the weight function is obtained by requiring that the second functional derivative of the excess free-energy
\begin{equation}\label{eq:c2}
	c^{(2)}(r_1,r_2;\rho)=-\frac{\beta\delta^2\mathcal{F}_{ex}[\rho]}{\delta\rho(r_1)\delta\rho(r_2)}
\end{equation}
\noindent gives an accurate description of the correlations of the homogeneous fluid. 

Different WDA functionals result from different approximations for the weight function. We use a modified version of the WDA proposed by Tarazona, developed by Kim and coworkers \cite{Kim}. In this approximation

\begin{equation}
	\tilde{\rho}(\vec{r})=\int d\vec{r'} \rho(\vec{r}')\omega(\vec{r}-\vec{r}',\rho_b),
\end{equation}

\noindent where $\rho_b$ is the bulk density. Following Tarazona, $\omega(\vec{r},\rho)$ is expanded in powers of $\rho$ in order to reduce the computational effort:

\begin{equation}
	\omega(r,\rho)=\omega_{0}(r)+\omega_{1}(r)\rho+\omega_{2}(r)\rho^2.
\end{equation}

\noindent The same is done for  $\bar{\rho}(\vec{r})$:

\begin{equation}\label{eq:bar_rho}
	\bar{\rho}(\vec{r})=\rho_0(\vec{r})+\rho_1(\vec{r})\tilde{\rho}(\vec{r})+\rho_2(\vec{r})\tilde{\rho}(\vec{r})^2,
\end{equation}

\noindent where

\begin{equation}\label{eq:tilde_rho}
	\tilde{\rho}(\vec{r})=\rho_0(\vec{r})+\rho_1(\vec{r})\rho_b+\rho_2(\vec{r})\rho_b^2,
\end{equation}

\noindent with

\begin{equation} \label{eq:rho_i}
	\rho_{i}(\vec{r})=\int d\vec{r}' \rho(\vec{r}')\omega_{i}(\vec{r}-\vec{r}')\qquad i=0,1,2.
\end{equation}

\noindent The grand potential functional

\begin{equation}\label{eq:grand_potential}
	\beta\Omega[\rho]=\beta F[\rho]+\beta\int d\vec{r}\, (V_{ext}(\vec{r})-\mu)\rho(\vec{r})
\end{equation}

\noindent is minimized with respect to variations of $\rho(\vec{r})$:

\begin{equation}
	\frac{\beta\delta \Omega[\rho]}{\delta \rho(\vec{r})}=\frac{\beta\delta F[\rho]}{\delta\rho(\vec{r})}-\beta(\mu-V_{ext}(\vec{r}))=0,
\end{equation}

\noindent which yields the equation for the equilibrium density profile, 

\begin{equation}
	\frac{\beta\delta F[\rho]}{\delta \rho(\vec{r})}= \beta\mu-\beta V_{ext}(\vec{r}).
\end{equation}

The first derivative of the excess free energy functional is the single-particle direct correlation function $c^{(1)}(\vec{r})$ and putting together Eqs.~\ref{eq:F_id},~\ref{eq:F_ex},~\ref{eq:mapping} and~\ref{eq:rho_bar} we obtain

\begin{equation}\label{eq:Relc1}
 	\beta\mu-\beta V_{ext}(\vec{r})=\ln\rho(\vec{r})-c^{(1)}[\vec{r};\rho],
\end{equation}

\noindent with

\begin{equation}\label{eq:c1}
	c^{(1)}[\vec{r};\rho]=-\beta f(\bar{\rho}(\vec{r}))-\beta\int d\vec{r}' \rho(\vec{r}') f'(\bar{\rho}(\vec{r}))\frac{\delta\bar{\rho}(\vec{r}')}{\delta \rho(\vec{r})},
\end{equation}

\noindent and 

\begin{eqnarray}
	\frac{\delta\bar{\rho}(\vec{r}')}{\delta \rho(\vec{r})}
	&=&\omega(\vec{r}-\vec{r}',\tilde{\rho}(\vec{r}'))\nonumber \\
	&+&\omega(\vec{r}-\vec{r}',\rho_b)
	\int d\vec{r}'' \rho(\vec{r}'') \omega'(\vec{r}'-\vec{r}'',\tilde{\rho}(\vec{r}')).\nonumber\\
\end{eqnarray}

The chemical potential $\mu$ is evaluated from the homogeneous version of Eq.~\ref{eq:c1}:

\begin{eqnarray}\label{eq:Hom_c1}
	\beta\mu &=& \ln\rho_b-c^{(1)}(\rho_b)\nonumber\\
	&=& \ln\rho_b-\beta f(\rho_b)-\beta\rho_b f'(\rho_b).
\end{eqnarray}

\noindent Finally, combining Eqs.~\ref{eq:Relc1} and \ref{eq:Hom_c1}, we obtain the density profile:

\begin{equation}
	\rho(\vec{r})=\rho_b\exp[-\beta V_{ext}(\vec{r})+c^{(1)}[\vec{r};\rho]-c^{(1)}(\rho_b)].
\end{equation}

The excess free energy is the sum of the free energy of HS given by the Carnahan-Starling approximation \cite{HansenMcDonald} and the bonding contribution given by Eq.~\ref{eq:Bond_Free_En}.

At a planar hard-wall, the external field is $V_{ext}=0$ for $z>0$ and infinite otherwise, and the density profile depends only on the distance $z$ from the wall:

\begin{eqnarray}\label{profile}
	\rho(z)&=&\rho_b \exp[c^{(1)}(z;[\rho])-c^{(1)}(\rho_b)]\qquad z>\sigma/2, \nonumber\\
	&=& 0 \qquad z < \sigma /2 .
\end{eqnarray}

The weight functions $\omega_i(\vec{r})$ for this system, defined in Ref.~\cite{Kim}, are:  

\begin{eqnarray}
	\omega_0(z) &=& \frac{3}{4\pi\sigma^3}, \quad |z|<\sigma,\nonumber\\
	& &\nonumber\\
	&=& 0 \quad \text{ otherwise},
\end{eqnarray}

\begin{eqnarray}
	\omega_1(z) &=& 0.21\sigma^2-1.49 z^2-0.18(\frac{z^2}{\sigma})^2+1.36\frac{|z|^3}{\sigma}, z<\sigma \nonumber\\
 	&=& -0.11\sigma^2 + 2.9 z^2+0.29(\frac{z^2}{s})^2-1.81\sigma |z|\nonumber\\
 	& & -1.6 \frac{z^3}{\sigma} \quad \sigma \leq z\leq 2\sigma, \nonumber\\	 
 	&=& 0  \quad \text{ otherwise},
\end{eqnarray}
and
\begin{eqnarray}
	\omega_2(z) &=& \frac{10\pi^2}{576} \sigma(\sigma^4-12(\sigma z)^2-5z^4 + 16\sigma |z|^3), |z|<\sigma, \nonumber\\
	& &\nonumber\\
	&=& 0 \quad \text{ otherwise}.
\end{eqnarray}

\section{Fundamental measure density functional}\label{FMT}

A second approach, uses a geometry-based density-functional to describe the excluded volume between spheres as well as the inhomogeneous associating free-energy. It was proposed by Yu and Wu \cite{yu:7094} to describe inhomogeneous mixtures of hard-spheres (HS) with an arbitrary set of interaction sites. In the following we consider a single-component fluid of HS with $n_p$ identical patches, in which case, the excess free-energy reads:
\begin{equation}
	\beta \mathcal{F}_{exc}[\rho(\vec r)]=\int d^3r \left\{\beta f_{HS}[n_\alpha(\vec r)]+\beta f_{bond}[n_\alpha(\vec r)]\right\},
\end{equation}
where $f_{HS}$ is the reduced excess free-energy of the fluid of HS, and $f_{bond}$ is the free-energy arising from the association of the particles. Both quantities depend on a set of weighted densities $\{n_\alpha(\vec r)\}$ (see later). 

\subsubsection{Hard sphere fluid}
The excluded volume interactions between HS are described by the Rosenfeld functional \cite{Rosenfeld}:

\begin{equation}
	\begin{aligned}
	\beta f_{HS} &= -n_0\ln(1-n_3)+\frac{n_1n_2-\vec n_{\upsilon1}\cdot\vec n_{\upsilon2}}{1-n_3} \\
	&+\frac{(n_2)^3-3n_2\vec n_{\upsilon2}\cdot\vec n_{\upsilon2}}{24\pi(1-n_3)^2}.
	\end{aligned}
\end{equation}

\noindent where we have dropped the spatial dependence of the weighted densities for convenience. $n_\alpha(\vec r)$ are convolutions of the density with the weight functions $w_\alpha(\vec r)$, which are related to geometrical properties of the particles:

\begin{equation}
	n_\alpha(\vec r)=w_\alpha(\vec r)*\rho(\vec r),
\end{equation}

\noindent where $*$ denotes the three-dimensional convolution $h(\vec r)*g(\vec r)=\int d^3x h(\vec x)g(\vec x-\vec r)$. For HS the weight functions are:

\begin{eqnarray}
	w_3(\vec r)=\Theta(R_S-|\vec r|),\\
	w_2(\vec r)=\delta(R_S-|\vec r|),\\
	w_1(\vec r)=w_2(\vec r)/(4\pi R_S),\\
	w_0(\vec r)=w_2(\vec r)/(4\pi R_S^2),\\
	\vec w_{\upsilon2}(\vec r)=w_2(\vec r)\frac{\vec r}{|\vec r|},\\
	\vec w_{\upsilon1}(\vec r)=\vec w_{\upsilon2}(\vec r)/(4\pi R_S).
\end{eqnarray}

\noindent $\delta(\cdot)$ is the Dirac-delta distribution and $\Theta(\cdot)$ is the Heaviside step function. $R_S=\sigma/2$ is the sphere radius.  
In planar geometry the one particle distribution function depends only on the normal distance from the wall, $z$. The weight functions are obtained integrating over the lateral coordinates,

\begin{equation}
	w_\alpha(z)=\int dx\int dy w_\alpha(\vec r).
\end{equation}

\noindent The resulting weight functions are 

\begin{eqnarray}
	w_3(z)=\pi(R_S^2-z^2)\Theta(R_S-|z|),\\
	w_2(z)=2\pi R_S\Theta(R_S-|z|),\\	
	w_1(z)=w_2(z)/(4\pi R_S),\\
	w_0(z)=w_2(4\pi R_S^2),\\
	\vec w_{\upsilon2}=2\pi z\Theta(R_S-|z|)\hat z,\\
	\vec w_{\upsilon1}=w_{\upsilon2}/(4\pi R_S),
\end{eqnarray}
 
\noindent with $\hat z$ the unit vector normal to the wall.

\subsubsection{Wertheim's inhomogeneous free-energy}

The bulk free energy of a fluid of particles with $n_p$ identical sites (eq.~\ref{eq:Bond_Free_En})
\cite{wertheim1,wertheim2,wertheim3,wertheim4} can be rewritten in terms of the bulk fraction of unbonded sites
$X_{b} \equiv 1-p_b$ as 

\begin{equation}\label{bulkfb}
	\beta f_{bond}=n_p\rho_{b}\left(\ln X_{b}-\frac {X_{b}}{2}+\frac{1}{2}\right)
\end{equation}   

\noindent with $\rho_{b}$ the bulk density, and $X_{b}$ the bulk fraction of unbonded sites, related to the thermodynamic variables through the law of mass action:

\begin{equation}
	X_{b}=\left(1+n_p\rho_{b}X_{b}\Delta\right)^{-1}.
\end{equation}
 
Yu and Wu \cite{yu:7094} generalized the bulk free energy, Eq. (\ref{bulkfb}), to inhomogeneous systems by including a new factor $\zeta=1-\vec n_{\upsilon2}\vec n_{\upsilon2}/n_2^2$ that incorporates the vectorial weight densities into the associating part of the free energy. The bonding free energy for inhomogeneous fluids reads:
\begin{equation}
	\beta f_{bond}[n_\alpha(\vec r)]=n_pn_0(\vec r)\zeta(\vec r)\left(\ln X(\vec r)-\frac{X(\vec r)}2+\frac12 \right).
\end{equation} 

\noindent where $X(\vec r)$ is the fraction of unbonded sites at position $\vec r$ given by the modified law of mass action:
\begin{equation}\label{eq:inhomog_X}
X(\vec r)=\left(1+n_pn_0(\vec r)\zeta(\vec r)X(\vec r)\Delta(\vec r)\right)^{-1}.
\end{equation}

The interaction between two sites determines $\Delta$. For the Kernel-Frenkel potential (where orientational and translational degrees of freedom are decoupled) $\Delta$ is given in terms of $g_{HS}(\vec r)$, the pair correlation function of the reference HS fluid, and $f_M$, the Mayer function:

\begin{equation}
	\Delta(\vec r)=\int d^3rg_{HS}(\vec r)f_M,\label{Delta}
\end{equation}

\noindent where $f_M=\exp(\beta \varepsilon)-1$, and the integral is over the bonding volume $v_b$. Following Yu and Wu \cite{yu:7094} we use a modified contact value for the pair correlation function:

\begin{equation}
	g_{HS}(\vec r)=\frac1{1-n_3}+\frac{\sigma n_2\zeta}{4(1-n_3)}+\frac{\sigma^2(n_2)^2\zeta}{72(1-n_3)^2}.
\end{equation}

\noindent Assuming that the pair correlation function is constant over the bonding volume, we approximate Eq. (\ref{Delta}) by:

\begin{equation}
	\Delta(\vec r)=v_b g_{HS}(\vec r)f_M.
\end{equation}

Finally, we minimize the grand potential,

\begin{equation}
	\Omega[\rho]=F[\rho]-\mu\int d^3r\rho(\vec r),
\end{equation}

\noindent to obtain the equilibrium density profiles. $\mu$ is the chemical potential and the integrals are computed using the trapezoidal rule and a step size $\Delta z=0.01\sigma$. 
We use a standard conjugated-gradients method to minimize $\Omega$. 

\bibliographystyle{apsrev4-1}

\end{document}